# Information Propagation Model in Hybrid Networks

Fuzhong Nian and Hongyuan Diao

*Abstract*—It is in practice impossible to describe the topology of a real network or its message propagation process using a single dynamic model. To address this issue, we constructed a new hybrid network model based on scale-free (SF), small-world (SW) features that functions as closely as possible to a real network. And the hybrid propagation model is constructed with susceptible-infected-susceptible (SIS), susceptible-infected-recovered (SIR) and susceptible-infected-recovered-susceptible (SIRS) model mixed in arbitrary proportions. The model applies the concepts of blockbuster effect and implicit node edges to reflect explosive spread as a significant characteristic of information propagation. A theoretical analysis and derivation of the new model in which hybrid networks were simulated revealed that the network degree distribution closely follows a power law. Using an improved similarity function to define the degree of closeness to real network cases, the proposed model was shown to be valid and very close to a real network.

## I. Information

INFORMATION spread has become an important sub-field of computational communication research as an inevitable outgrowth of the need to understand how information transmission has been changed by online social networks (OSNs). Information spread has been analyzed in various fields by a number of researchers [1-8] using two general approaches involving the analysis of complex network topologies and dynamic network models, respectively. The first approach [9-11] has a long history starting with regular scale-free (SF) networks and leading to the development of complex network models [12-15]. Using the latter approach, the conventional epidemic model has been applied to the modeling of rumor dynamics in OSNs [16]. The epidemic model has three classic models—the susceptible-infected-recovered (SIR) [17, 18], susceptible-infected-susceptible (SIS) [19-21], and susceptible-infected-recovered-susceptible (SIRS) [22, 23] models. The representative rumor propagation model (known as the DK model), which was proposed by Daley and Kendall, analyzes rumors using a random process method in which the audience is divided into three categories according to the rumor propagation effect [24].

More recently, Nian *et al.* proposed a new multi-relationship network with SF and SW features [25]. Chakraborty *et al.* analyzed the community network structure in large-scale production networks [26], Dai *et al.* proposed a weighted SF tree network [27], and Türker *et al.* introduced network models with Poisson-based edge location strategies that produced cluster-free and SF topologies [28]. Konstantin's work on clustering scale-free networks produced various network structures in which SF and SW features were combined [29, 30]. Sanatka *et al.* attempted to more accurately describe dynamic switching networks [31]. Many of the studies cited above addressed the issue of network reconfiguration to produce, for example, networks that combined into a community and clustered SF networks. Although such reconstructed models have a number of advantages, they face some unresolved problems. Realistic network relationships have at least two important relationship features that are characterized by either SW or SF networks. Examples of these include the typical friend relationship (SW) and actor cooperative (SF) networks, both of which occur in OSNs. However, it is difficult to define a hybrid network model with specific proportions in a manner that reflects real network complexity. To address this, new network models that mix SW and SF features in multiple ways and in varying proportions have been proposed.

The study of propagation dynamics has also received a high degree of attention in recent years. There are differences between the OSNs information propagation and virus transmission. Wang *et al.* developed an edge-based SIR model on a degree-dependent network [32], while Yang *et al.* established a rumor truth competition model on a two-layer network [33]. Lebensztayn *et al.* considered variations in the model [34], while Sumith *et al.* improved the model to develop the ReSIR rumor model [35]. Zhang *et al.* revealed that adaptive behavior during disease epidemics induces a more complex dynamic virus propagation model [36]. Chen *et al.* proposed a two-stage hybrid SIR model [37]. Although these studies have helped to improve the classical virus dynamics model, few of indeterminate number of multiple classical propagation models are considered. Under classical SIS models, the vertexes forget messages following reception. By contrast, under the SIR model vertexes do not forget messages and propagate them following reception, while, under the SIRS model, vertexes retain and propagate information immediately upon reception but then gradually forget it. A variety of models

This research is supported by the National Natural Science Foundation of China (No. 61263019; No.61863025), Program for International S&T Cooperation Projects of Gansu province (No.144WCGA166), Program for Longyuan Young Innovation Talents and the Doctoral Foundation of LUT.

Fuzhong Nian is with the School of Computer and Communication, Lanzhou University of Technology, LanZhou, 730050 China (e-mail: gdnfz@lut.cn).

Hongyuan Diao is with the School of Computer and Communication, Lanzhou University of Technology, LanZhou, 730050 China.

more accurately reflect the actual OSN information dissemination process, in which not all vertexes will adhere to a single classical propagation model. Consequently, hybrid propagation models have been designed to combine the SIS, SIR, and SIRS approaches in multiple ratios.

The phenomenon of Human Flesh Search (HFS), which originates in China [38-43], refers to a purpose-driven form of special information dissemination. To date, HFS research has primarily involved qualitative sociological analyses undertaken in China [38, 41, 43], and there has been only a limited amount of quantitative analytical HFS research from a science and engineering perspective [44]. Li et al. used the results of HFS searches to study the hot factors of evolutionary games [45], while Tan et al. reconstructed networks by adding links identified using a predictive link algorithm [46]. Kandhway et al. used node centrality and optimal control to vast Information spread [47]. Wang et al. have a study of the impact of uncertainty on behavior diffusion in social networks [48]. Gomez-Gardenes et al. studied local synergies associated with acquaintances [49]. The authors of these studies also looked at a number of influencing factors in the process of information dissemination [50]. Not all information in an OSN propagates at a large scale, and the degree of propagation is significantly affected by time and density of infection—two factors that can be modeled together using the blockbuster effect function. In this paper, we explore the blockbuster effect in the context of hybrid networks and hybrid propagation models.

The main contributions of this paper are as follows: (1) the hybrid information propagation model is presented based on the hybrid network model, hybrid propagation model and the blockbuster effect in Section 2; (2) the network attributes and transmission characteristics are depicted via mathematical analysis in Section 3; (3) the network structure and propagation process is simulated in Section 4; (4) the comparative analysis with the network data is described and the similarity function is defined to prove rationality of the constructed model.

## II. HYBRID INFORMATION PROPAGATION MODEL

### A) The Hybrid Network Model

Hybrid network models are built to reflect both SW and SF features. The first and second and third hybrid network are named network I, II and III, respectively.

Fig. 1(a) shows the network I analyzed in this paper. The percentage $a$ on the axis at the top of the figure gives the ratio of SF to total nodes, while $(1 - a)$ gives the ratio of SW to total nodes. The red dots and lines indicate SF nodes and linkages, respectively, and reflect the degree of SF connectedness of the network. Similarly, the blue dots and lines indicate SW nodes and linkages, respectively, reflecting the SW characteristics of the network.

**Algorithm 1:** Constructing the network I.

*(SW characteristics)*
1. Initializing the rule graph: Define an annular nearest neighbor-coupled network with a total of $(1-a)N$, each of which is connected to a total of K/2 neighbor nodes, where K is an even number.
2. Randomization reconnection: With a probability p, arbitrarily rewire each edge by leaving one vertex of the edge unchanged while arbitrarily connecting to another vertex in the network. Multi-edges and self-looping must be avoided in the process.

*(SF characteristics)*
3. Following the preferential attachment principle, add aN SF nodes in turn. The relationship between the degree $k_i$ of node i and the probability $\Pi_i$ that a new node is connected to an existing node i is given by $\Pi_i = \frac{k_i}{\sum_j k_j}$.

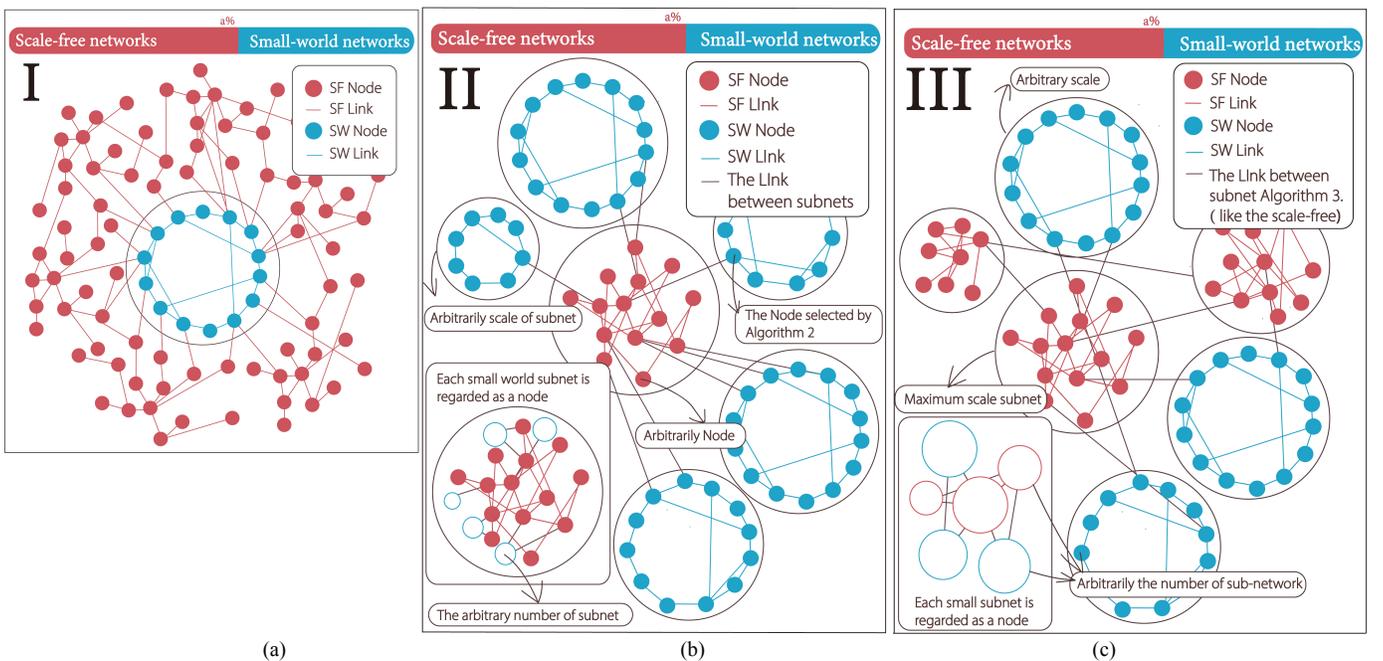

Fig. 1 Schematic showing the network I, II and III connection principle.

Fig. 1(b) shows the network II introduced in this paper. As in Fig. 1(a), $a$ and $(1-a)$ on the top horizontal axis are the ratios of SF and SW nodes, respectively, to the total number of nodes. Similarly, SF and SW nodes and links are colored red and blue, respectively. The links between subnets (dark brown lines), which are generated using *Algorithm 2*, indicate where random selected nodes in SF subnets are connected with random selected nodes in SW subnets. The total number of nodes in each SW subnet is arbitrarily determined to lie between 20–100% of the overall network node count based on a node selection probability of 0.5. Both the number of SW subnets and the number of subnet nodes are arbitrarily selected. Each SW subnet is regarded as a node and connected to an SF network.

**Algorithm 2:** Construction of the network II.
*(SF characteristics)*
1. Generate a fully connected network with three nodes
2. Following the preferential attachment principle, add ($aN$-3) SF nodes in turn. The relationship between the degree $k_i$ of node $i$ and the probability $\Pi_i$ that a new node is connected to an existing node $i$ is given by $\Pi_i = \frac{k_i}{\Sigma_j k_j}$。

*(SW characteristics)*
3. Arbitrarily generate the number of SW subnets, $U_{2WS}$, and the number of nodes in each SW subnet, $P_i$, which fulfills $(1\text{-}a)N = \sum_i^U P_i$.
4. Generate the $U_{2WS}$ small-world subnets in turn based on the following rule map: given an annular nearest neighbor coupling network with $P_i$ nodes, connect each nodes to a total of $K/2$ neighbor nodes, where $K$ is an even number.
5. Randomization reconnection: With a probability $p$, arbitrarily rewire each edge by leaving one vertex of the edge unchanged while arbitrarily connecting to another vertex in the network. Multi-edges and self-looping must be avoided in the process.
6. Choose an arbitrary number $\xi_L \in \left[1, 10\% \times P_i\right]$ of nodes from each SW subnet. Place the i-th arbitrarily selected node in a list, $L_i$, of nodes connected to arbitrarily selected nodes in SF networks.

Fig. 1(c) shows the network III discussed in this paper. The top axis and red and blue nodes indicate the same features as in Figs. 1(a) and 1(b), while the links between subnets (dark brown lines) represent connections that are constructed in accordance with the SF method applied in *Algorithm 3*. Each new subnet is connected to $m$ existing nodes with a connection probability that satisfies the condition that subnets with more nodes are more likely to be selected. This is done by selecting the nodes based on a weighting that is positively correlated with the node degree.

**Algorithm 3:** Constructing the network III.
The network III contains connections in which entire subnets serve as nodes. There are a total of $U$ subnets, with $U_{WS}$ SW feature subnets and $U_{BA}$ SF feature subnets, where $U = U_{BA} + U_{WS}$. The network is constructed as follows:
1. Arbitrarily generate the number of subnets, $U$.
2. Set the number of nodes with SW and SF features as $(1\text{-}a)N$ and $aN$, respectively, and define the network for each subnet in turn so that $U = U_{BA} + U_{WS}$.
3. Assign $\varphi_{I_{WS}}$ nodes to the $I_{WS}$th subnet and $\varphi_{I_{BA}}$ to the $I_{BA}$-th subnet so that $(1\text{-}a)N = \sum_{I_{WS}}^{U_{WS}} I_{WS} \times \varphi_{I_{WS}}$ and $aN = \sum_{I_{BA}}^{U_{BA}} I_{BA} \times \varphi_{I_{BA}}$i
4. If the number of subnet nodes of the t-th subnet is $\varphi_i < 4$, the subnet is s a fully connected network.

*(SW characteristics)*
5. Generate $U_{WS}$ SW subnets in turn using the following rule map: given an annular nearest neighbor coupling network with $\varphi_{I_{WS}}$ nodes, connect each nodes to a total of $K/2$ neighbor nodes, where $K$ is an even number.
6. Randomization reconnection: With a probability $p$, arbitrarily rewire each edge by leaving one vertex of the edge unchanged while arbitrarily connecting to another vertex in the network. Multi-edges and self-looping must be avoided in the process.

*(SF characteristics)*
7. Generate $U_{BA}$ SW subnets in turn to produce a fully connected network with three nodes.
8. Following the preferential attachment principle, add the ($\varphi_{I_{BA}}$-3) SF nodes in turn. The relationship between the degree $k_i$ of node $i$ and the probability $\Pi_i$ that a new node is connected to $i$ is given by $\Pi_i = \frac{k_i}{\Sigma_j k_j}$。
9. Treat all subnets as a node-generation network:
    a) Arbitrarily select the $E$-th subnet.
    b) Connect the other subnets in turn to the existing subnets. Each new subnet is connected to an existing subnet through two edges with a probability of $\varpi_I = \frac{\varphi_I}{\Sigma_q \varphi_I}$, where $\varphi_I$ is the summary point of subnet $E$.
    c) Connect the two selected two subnets at a node selected with probability $Q_i = \frac{k_i}{\Sigma_j k_j}$, where $k_i$ is the degree of node $i$.

*B) Hybrid Propagation Model*
Actual human activity across a real network is diverse, making it difficult to completely describe how information propagates in a crowd using only one propagation model. We therefore propose that an accurate network requires the coexistence of multiple dynamic propagation models. Under our proposed hybrid propagation mode, nodes can be in three possible states: ignorant, spreader, and stifler. A node that is "ignorant" with respect to a specific message has not received it. A "spreader" is a node that has obtained the message and propagates it, while a "stifler" receives the message but does not propagate it. Spreaders propagate messages to their neighbor nodes, which in turn transform into spreaders with an effective propagation probability $\lambda$. Over time, a fraction $u$ of the spreaders ($I$) will transform into ignorant ($S$) nodes with a forgetting probability $\beta$; such nodes will be of the SIS propagation type. Similarly, a fraction $w + q$ of the spreaders ($I$) transform into stiflers ($R$) with a recovery probability $\beta$; of these, a fraction ($q$) will be of the SIR type and will not change again, while another fraction ($w$) will convert back to ignorant ($S$) with probability $\sigma$ and be characterized by the SIRS model type. Thus, the effective propagation rate of the model is $\lambda = \beta/\gamma$, where $u + w + q = 1$.

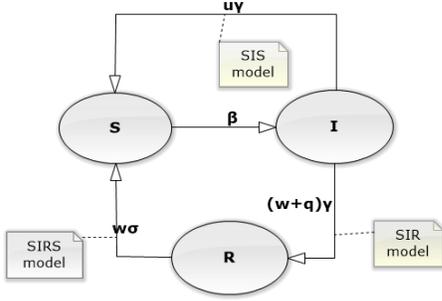

Fig. 2 Schematic of model changing under hybrid propagation network.

*C) The Blockbuster Effect*

People often overlook potential relationships, including friends whom they have not contacted with in a long time or accounts that they do not frequently log into. Such relationships can be discovered by applying the HFS process. If an individual is represented by a vertex and a contact is represented by an edge, then the common relationships between individuals can be defined as dominant edges while potential relationships can be defined as implicit edges. In this manner, a network of average degree $\alpha$ is established in which $\delta$ edges leading into each node are dominant in the network while the rest are implicit. In this case, $\alpha > \delta > 1$. The corresponding network is shown in Fig. 3.

As an example of this network in operation, we take the case of a message that has spread to a certain scale and has attracted a high degree of attention. The spreaders in the network will circulate the message extensively in a number of hidden ways. As the message spreads throughout the crowd, it will be transmitted to broader scales in increasingly shorter times, making it likelier for the blockbuster effect to occur. In this process, the blockbuster effect changes with time in a manner that is positively correlated with the infection density but negatively correlated with time. Consequently, the blockbuster effect is defined as follows:

$$\Phi(t) = \frac{i(t)}{lg(t+1)}, \quad (1)$$

where $i(t)$ is the proportion of spreaders in the total crowd at time t. The value of the blockbuster effect in propagation selects a critical point $\varphi$ based on the trigger decision function as follow:

$$\Gamma(t) = \begin{cases} 1 & \Phi(t) \geq \varphi \\ 0 & \Phi(t) < \varphi \end{cases} \quad (t \leq \frac{T}{2}), \quad (2)$$

where T is the total number of propagation rounds in the experiment. When $t \leq \frac{T}{2}$, $\Gamma(t) = 0$ and $\Phi(t) \geq \varphi$, the blockbuster effect is triggered and $\Gamma(t) = 1$.

As information propagates through the network, implicit edges are transformed into dominant edges, as indicated by the process shown in right-hand side of Fig. 3. As long as $\Phi(t) < \varphi$, however, the blockbuster effect is not triggered, $\Gamma(t) = 0$ and $< k > = \delta$. In this case, few implicit edges have changed, and the network is better characterized by the left-hand side of Fig. 3, when $\Gamma(t) = 1$ and $< k > = \alpha$.

The blockbuster effect is touched off when the decision function $\Gamma(t)$ is triggered at $t > \frac{T}{2}$:

$$\Gamma(t) = \Gamma(\frac{T}{2}) \quad (t > \frac{T}{2}), \quad (3)$$

$$< k > = \begin{cases} \alpha & \Gamma = 1 \\ \delta & \Gamma = 0 \end{cases}. \quad (4)$$

In this state, the rate of spread as a result of the transformation of implicit edges to dominant edges has reached a critical average value.

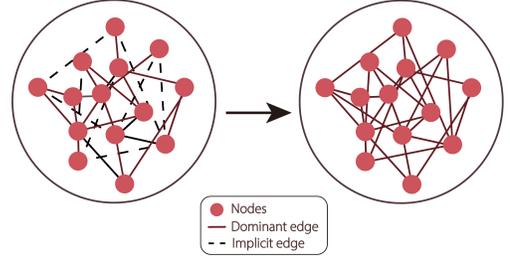

Fig. 3 Schematic of blockbuster effect triggering.

### III. MATHEMATICAL ANALYSIS

In this section, we perform an analysis of the degree of distribution in the three hybrid networks. To do this, we establish a network with WS feature components. If the reconnecting probability of the network $p = 0$, the degree of each node is $K$ (even); if $p > 0$, the algorithm of randomization reconnection rules is implemented using the SW model, and each node is still connected in the clockwise direction to at least the $K/2$ original edges in that direction. In simpler terms, the degree of each node is at least $K/2$, and therefore the degree of node $i$ is $k_i = s_i + K/2$, where $s_i \geq 0$ is an integer [51].

The parameter $s_i$ is further divided into two parts unchanged, $s_i = s_i^1 + s_i^2$, where $s_i^1$ is the number of edges, out of the original $K/2$ edges connected to node $i$ in the counterclockwise direction. In this case, the probability that each edge remains unchanged is $1 - p$. The parameter $s_i^2$ indicates the number of long-range edges connected to node $i$ through the randomization reconnection mechanism, which occurs with a probability $p/N$. These parameters are derived analytically as follows:

$$P_1(s_i^1) = \binom{\frac{K}{2}}{s_i^1}(1-p)^{s_i^1} p^{\frac{K}{2}-s_i^1}, \quad (5)$$

$$P_2(s_i^2) \simeq \frac{\left(\frac{pK}{2aN}\right)^{s_i^2}}{(s_i^2)!} e^{-\frac{pK}{2aN}}, \quad (6)$$

for any node with degree $k \geq \frac{K}{2}$, $s_i^1 \in \left[0, min\left(k - \frac{K}{2}, K/2\right)\right]$. Therefore, when $k \geq K/2$,

$$P_s(k) = \sum_{n=0}^{min\left(k-\frac{K}{2}, \frac{K}{2}\right)} \binom{K/2}{n}(1-p)^n p^{\frac{K}{2}-n} \frac{\left(\frac{pK}{2aN}\right)^{k-\frac{K}{2}-n}}{\left(k - \frac{K}{2} - n\right)!} e^{-\frac{pK}{2aN}}. \quad (7)$$

Thus, when $k < K/2$, $P_s(k) = 0$.

The major theoretical approaches currently used to understand the distribution in a Barabási–Albert (BA) SF network include continuum theory [52], the master equation method [53], and mean-field theory [54]. The asymptotic results obtained by the three methods are similar.

Here, we analyze a network with $aN$ nodes, in which we note that the degree of node $i$ at time $t$ is $k_i(t), t \in [aN, N]$. We can analyze the SF network characteristics using mean-field theory. Because $aN$ edges are ignored in the initial network, the network has $aN + t \approx t$ nodes at time $t$. With the addition of each new node, the probability of the degree of existing node $i$ changing (increasing by $1$) is

$$m\Pi_i = \frac{mk_i(t)}{\sum_{j=1}^{aN+t} k_j(t)} \approx \frac{mk_i(t)}{2mt} = \frac{k_i(t)}{2t}. \quad (8)$$

We use the average field theory to approximately determine the degree distribution of a network with BA SF features [54] under the following continuity assumptions: that the time, $t$, is not discrete but continuous; and that the node degree is not an integer but an arbitrary real number. Under these two assumptions, (8) can be interpreted as the rate of change of degree of node $i$ and the evolution of the network can be approximated using the mean field equation of single node evolution:

$$\frac{\partial k_i(t)}{\partial t} = \frac{k_i(t)}{2t}. \quad (9)$$

For a node $i$ that joins the network at time $t_i$, the initial condition of (9) is $k(t_i) = m$, and the node evolution is given by

$$k_i(t) = m\left(\frac{t}{t_i}\right)^{\frac{1}{2}}. \quad (10)$$

As $t \to \infty$, the degree distribution $P_b(k(t))$ converges to the steady-state index distribution $P_b(k)$ defined by the probability

$$P_b(k) = \frac{\partial P_b\left(k_i(t) > \frac{m^2 t}{k^2}\right)}{\partial k}. \quad (11)$$

Combining (10) and (11) gives

$$P_b(k_i(t) < k) = P_b\left(t_i > \frac{m^2 t}{k^2}\right), \quad (12)$$

and therefore

$$P_b\left(t_i > \frac{m^2 t}{k^2}\right) = 1 - P_b\left(t_i \leq \frac{m^2 t}{k^2}\right) = 1 - \frac{m^2 t}{k^2(aN + t)}. \quad (13)$$

Substituting (13) into (11) with $t = (1-a)N$ gives

$$P_b(k) = \frac{\partial P_b(k_i(t) < k)}{\partial k} = 2m^2 \frac{t}{aN + t} \frac{1}{k^3}$$

$$\approx 2m^2(1-a)\frac{1}{k^3}. \quad (14)$$

Under the network I, there are a total of $N$ shared nodes, comprising $aN$ SF nodes and $(1-a)N$ SW nodes. The average node degrees of the SW and SF networks are $<k>_s = 2K$ and $<k>_b = 2m$, respectively, where $K = m$. The power-law distribution function of the hybrid network and its average degree are given by

$$P(k) = \begin{cases} P_s(k) & [1, m) \\ P_s(k) + 2m^2(1-a)\frac{1}{k^3} & [m, +\infty) \end{cases}, \quad (15)$$

and

$$<k> = K + \left\lfloor \frac{K}{2} \right\rfloor, \quad (16)$$

respectively.

Correspondingly, under the network II and III. The power-law distribution function of the hybrid network is given by

$$P(k) = \begin{cases} P_s(k) & [1, m) \\ 2m^2(1-a)\frac{1}{k^3}\left[P_s(k) + 2m^2(1-a)\frac{1}{k^3}\right] & [m, +\infty) \end{cases}.$$

The hybrid propagation model for this hybrid network model is described as follows. We construct a hybrid network that satisfies the features of growth incorporation, preferential attachment, and SW characteristics. After a sufficient number of rounds, we iteratively obtain a network with $N$ nodes, which has a degree distribution as described in (15) and an average degree as described in (16). Among the nodes of degree $k$, the proportions of ignorant, spreader, and stifler nodes are $s_k(t)$, $i_k(t)$, and $r_k(t)$, respectively. The average field dynamic equations of the SF network are then given as follows:

$$\begin{cases} \frac{ds_k(t)}{dt} = -\lambda k s_k(t)\Theta(t) + u i_k(t) + w\sigma r_k(t) \\ \frac{di_k(t)}{dt} = \lambda k s_k(t)\Theta(t) - i_k(t) \\ \frac{dr_k(t)}{dt} = (w+q)i_k(t) - w\sigma r_k(t) \end{cases}, \quad (17)$$

In (17), where $u + w + q = 1$. The first term on the right-hand side of the first equation indicates the density of new known nodes with degree $k$ generated by propagation from existing spreaders, with $\Theta(t)$ indicating the probability that any nodes are connected to a given vertex. The second term on the right-hand side of the first equation indicates the probability $u$ that a node with degree $k$ belongs to the SI model. The third term on the right-hand side of the first equation indicates that stiflers are converted to spreaders with probability $w\sigma$.

The probability that a given edge in the SF network is linked to a spreader node is given by

$$\Theta(t) = \frac{\sum_k k P(k) i_k(t)}{\sum_s s P(s)} = \frac{1}{<k>} \sum_k k P(k) i_k(t). \quad (18)$$

For each $k$, $s_k(t)$, $i_k(t)$, and $r_k(t)$ all satisfy the standardized condition

$$s_k(t) + i_k(t) + r_k(t) = 1. \quad (19)$$

The static condition in which no messages propagate in the network is given by

$\frac{ds_k(t)}{dt} = 0, \frac{di_k(t)}{dt} = 0, \frac{dr_k(t)}{dt} = 0$, and $i_k(t) = 0$,

which, from (17), becomes

$$-\lambda k s_k(t)\Theta(t) + u i_k(t) + w\sigma r_k(t) = 0 \qquad (20)$$

Combining (19) and (20) gives

$$s_k(t) = \frac{(u - w\sigma)i_k(t) + w\sigma}{\lambda k \Theta(t) + w\sigma} \qquad (21)$$

From the second line of (17), we get

$$s_k(t) = \frac{(w + q)i_k(t) + u i_k(t)}{\lambda k \Theta(t)}, \qquad (22)$$

which, because $w + q + u = 1$, reduces to

$$s_k(t) = \frac{i_k(t)}{\lambda k \Theta(t)} \qquad (23)$$

Substituting (23) into (21) gives

$$i_k(t) = \frac{\lambda w\sigma k \Theta(t)}{\lambda(1 - u + w\sigma)k\Theta(t) + w\sigma} \qquad (24)$$

From the above, it is obvious that the solutions $i_k(t) = 0$ and $\Theta(t) = 0$ are always satisfied with (24): without obtaining a non-zero static solution ($i_k(t) \neq 0$), the right- and left-hand sides of (24) can be represented as a function $F(\Theta)$ in $\Theta$, where $0 < \Theta \leq 1$. If this formulation has a non-trivial solution, it must satisfy the following:

$$\frac{dF(\Theta)}{d\Theta}\Big|_{\Theta=0} \geq 1, \qquad (25)$$

which is equivalent to

$$\frac{d}{d\Theta}\left\{\frac{1}{<k>}\sum_k kP(k)\left\{\frac{\lambda w\sigma k\Theta(t)}{\lambda(1-u+w\sigma)k\Theta(t)+w\sigma}\right\}\right\}\Big|_{\Theta=0} \qquad (26)$$

$$= \frac{\lambda}{<k>}\sum_k k^2 P(k) w\sigma \geq 1$$

For a hybrid network,

$$P(k) = \begin{cases} P_s(k) & [1, m) \\ P_s(k) + 2m^2\dfrac{(1-a)N}{N}\dfrac{1}{k^3} & [m, +\infty) \end{cases} \qquad (27)$$

When the threshold in (27) is equivalent to that in (26),

$$\frac{\lambda_c}{<k>}\sum_k k^2 P(k) w\sigma = 1$$

$$\lambda_c = \frac{<k>}{<k>^2 w\sigma}, \qquad (28)$$

at which time the following holds:

$$<k> = \sum_k kP(k)$$

$$= \begin{cases} \sum_k kP_s(k) & [1, m) \\ \sum_k kP_s(k) + \sum_k 2m^2(1-a)\dfrac{1}{k^2} & [m, +\infty) \end{cases}$$

$$<k>^2 = \sum_k k^2 P(k)$$

$$= \begin{cases} \sum_k k^2 P_s(k) & [1, m) \\ \sum_k k^2 P_s(k) + \sum_k 2m^2(1-a)\dfrac{1}{k} & [m, +\infty) \end{cases}$$

Combining the two formulations above into (28) gives

$$\lambda_c = \begin{cases} \dfrac{\sum_k kP_s(k)}{w\sigma \sum_k k^2 P_s(k)} & k \in [1, m) \\[2mm] \dfrac{\sum_k kP_s(k) + 2m^2(1-a)\sum_k \dfrac{1}{k^2}}{\left\{\sum_k k^2 P_s(k) + 2m^2(1-a)\sum_k \dfrac{1}{k}\right\}w\sigma} & k \in [m, +\infty). \end{cases} \qquad (29)$$

The approximations $\sum_k \frac{1}{k^2}$ and $\sum_k \frac{1}{k}$ can be calculated using the continuous similarity of $k$ as

$$\sum_k \frac{1}{k^2} \to \int_m^M \frac{1}{k^2} dk = \frac{1}{m} - \frac{1}{M} = \frac{M - m}{mM}, \qquad (30)$$

$$\sum_k \frac{1}{k} \to \int_m^M \frac{1}{k} dk = \ln\frac{M}{m}, \qquad (31)$$

where the maximum value of the degree of a node is $M$. Substituting (30) and (31) into (29) gives

$$\lambda_c \approx \begin{cases} \dfrac{\sum_k kP_s(k)}{w\sigma \sum_k k^2 P_s(k)} & k \in [1, m) \\[2mm] \dfrac{\sum_k kP_s(k) + 2m(1-a)\dfrac{M-m}{M}}{\left\{\sum_k k^2 P_s(k) + 2m^2(1-a)\ln\dfrac{M}{m}\right\}w\sigma} & k \in [m, +\infty). \end{cases} \qquad (32)$$

If $M$ is sufficiently large (i.e., if $m$ is constant, $M$ approaches $N$), $M - m \approx M$ and

$$\frac{1}{\lambda_c} \approx \begin{cases} \dfrac{w\sigma \sum_k k^2 P_s(k)}{\sum_k kP_s(k)} & k \in [1, m) \\[2mm] \dfrac{\left\{\sum_k k^2 P_s(k) + 2m^2(1-a)\ln\dfrac{M}{m}\right\}w\sigma}{\sum_k kP_s(k) + 2m(1-a)} & k \in [m, +\infty), \end{cases} \qquad (33)$$

and

$$\frac{1}{\lambda_c} \approx \begin{cases} \dfrac{w\sigma \sum_k k^2 P_s(k)}{\sum_k kP_s(k)} & k \in [1, m) \\[2mm] w\sigma\left[\dfrac{\sum_k k^2 P_s(k)}{\sum_k kP_s(k) + 2m(1-a)} + \dfrac{2m^2(1-a)\ln\dfrac{M}{m}}{\sum_k kP_s(k) + 2m(1-a)}\right] & k \in [m, +\infty) \end{cases}$$

$$\approx \begin{cases} \dfrac{w\sigma \sum_k k^2 P_s(k)}{\sum_k kP_s(k)} & k \in [1, m) \\[2mm] w\sigma\left[\dfrac{1}{\dfrac{\sum_k kP_s(k)}{\sum_k k^2 P_s(k)} + \dfrac{2m(1-a)}{\sum_k kP_s(k)}\sum_k kP_s(k)} + \dfrac{m\ln\dfrac{M}{m}}{\dfrac{\sum_k kP_s(k)}{2m(1-a)} + 1}\right] & k \in [m, +\infty) \end{cases},$$

where $\sum_k kP_s(k)$ and $\sum_k k^2 P_s(k)$ are constant values in the formula, and therefore

$$\Upsilon \approx \frac{\sum_k k^2 P_s(k)}{\sum_k kP_s(k)}, \quad \Psi = \frac{m(1-a)}{\sum_k kP_s(k)}.$$

From (33), we obtain

$$\frac{1}{\lambda_c} \approx \begin{cases} w\sigma Y & k \in [1, m) \\ w\sigma \left( \dfrac{Y + 1 + 2\Psi \, m\ln\dfrac{M}{m}}{1 + 2\Psi} \right) & k \in [m, +\infty) \end{cases} \quad .(34)$$

1) $\lambda_c$ increases as $\sigma$ decreases.

Increasing the probability of stiflers ($R$) in the OSN means that messages are less likely spread through the crowd and the propagation threshold is increased. Reducing the probability of stiflers ($R$), on the other hand, increases the number of nodes that are transformed into unknown or spreader states, making it easier for the message to spread through the crowd and reducing the propagation threshold.

2) $\lambda_c$ increases as $w$ decreases.

The parameter $w$ is the probability that a node will transform into SIR mode. As the penetration of SIR nodes increases in a network, the probability of message spread will be reduced and the propagation threshold will increase. Conversely, as the number of SIR nodes decreases more nodes will convert to the SIS and SIRS models, which are more conducive to a wider spreading of messages than the SIR model [18, 19, 23]. Thus, if $w$ decreases, the propagation threshold will be reduced.

3) $\lambda_c$ decreases as $Y$ decreases.

Increasing $Y$ to cause more people (nodes) have extensive connections (higher degrees) will enhance message spread and reduce the propagation threshold. Conversely, reducing $Y$ will result in fewer extensive connections (lower degrees) and hinder the spreading of information, increasing the propagation threshold.

## IV. SIMULATION RESULTS

*A) Simulation of the Degree Distributions within the three Hybrid Network Models*

A power-law-like degree distribution map of the network I, corresponding to the results of *Algorithm 1*, is shown in Fig. 4. In the graph, the distribution trends of the model at three different hybrid network ratios are compared. It is seen that, when the ratios of SW and SF nodes to the total number of nodes are 0.99 and 0.01, respectively, the curve (square symbols) is dominated by a Poisson distribution below $k=10$. Above $k=10$, the distribution gradually changes to that of a power-law. At SW and SF fractions of 0.80 and 0.20, respectively (circles), a Poisson distribution dominates below $k= 10^{0.7}$, after which a power-law distribution gradually appears. At SW and SF node proportions of 0.99 and 0.01, respectively (triangles), a Poisson distribution dominates below $k = 10^{0.3}$ and is gradually replaced by a power-law distribution thereafter. Below $k = 10^{0.4}$, both the first (0.99/0.01) and second (0.8/0.2) curves have high degree distributions because, in both cases, the number of nodes in the SW network is above a certain size. As a result of the hybrid network structure, the networks, which have identical average degree and reconnection rates, both converge on power-law distributions as the degree increases but with somewhat inconsistent forms. The third (0.01/0.99) curve diverges significantly from the other two curves below $k=10^{0.4}$ and does not follow a Poisson distribution. This follows primarily from the fact the network producing the third curve has too few SW network nodes. Above $k = 10^{0.4}$, the orders of all three curves begin to change, with all three converging on power-law-like distributions above $k=10^{0.7}$. Increasing the proportion of SW nodes within the network causes the degree distribution to approach that of a Poisson distribution, while increasing the proportion of SF nodes moves the degree distribution toward a power-law-like distribution.

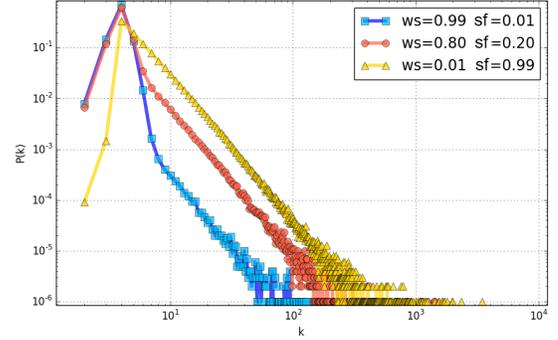

Fig. 4 Power-law-like distributions produced by simulation of the network I. N=1,000,000.

A degree distribution map comparing the results obtained simulating the network II using *Algorithm 2* at three hybrid network ratios is shown in Fig. 5. It is seen that, with a SW/SF node mix of 0.99/0.01 (square symbols), the curve is dominated by a Poisson distribution trend below $k=10$ and, above $k=10$, a power-law distribution in which there is a rapid fluctuation with the degree increasing and then decreasing. At SW and SF fractions of 0.80 and 0.20, respectively (circles), a Poisson distribution dominates below $k=10^{0.7}$, after which a power-law distribution gradually appears. Finally, SW and SF node proportions of 0.01 and 0.99, respectively (triangles), a Poisson distribution dominates below $k = 10^{0.3}$ and is gradually replaced by a power-law distribution thereafter. As in the case of the network I, increasing the proportion of SW nodes leads to an increasingly Poisson distribution, while increasing the number of SF nodes leads to an increasingly power-law distribution.

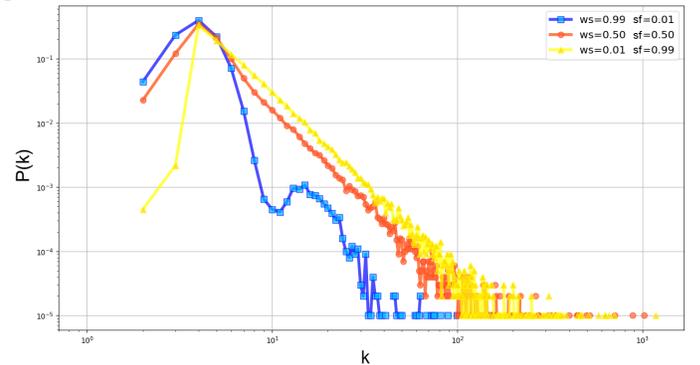

Fig. 5 Power-law-like distributions produced by simulation of the network II. N = 1,000.

The degree distribution map of the network III, constructed by *Algorithm 3*, is shown in Fig. 6. At a SW/SF node ratio of 0.99/0.01 (squares), the curve is dominated by a Poisson distribution trend below $k =10$, which falls off steeply thereafter. This occurs primarily because the network in this case is a hybrid that is so small that the algorithm produces a high proportion of nodes with the SF feature, all of which form small, fully connected networks. In the SW/SF 0.80/0.20 and 0.01/0.99 networks, however (circles and triangles, respectively) the Poisson-to-power-law transitions occur at $k = 10^{0.7}$ and $k = 10^{0.3}$, respectively. As in the network I and II cases, increasing the proportion of SW nodes within the network causes the degree distribution to approach that of a Poisson distribution, while increasing the proportion of SF nodes moves the degree distribution toward a power-law distribution.

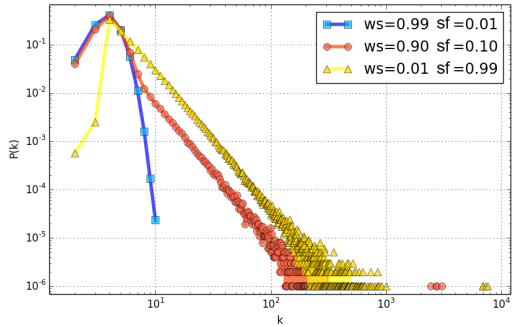

Fig. 6 Power-law-like distributions produced by simulation of the network III. N = 1,000,000.

B) *Hybrid Network Model Simulation*

We then performed multiple simulations of the three hybrid network models, with the results plotted in Fig. 7 (network I), Fig. 8 (network II), and Fig. 9 (network III).

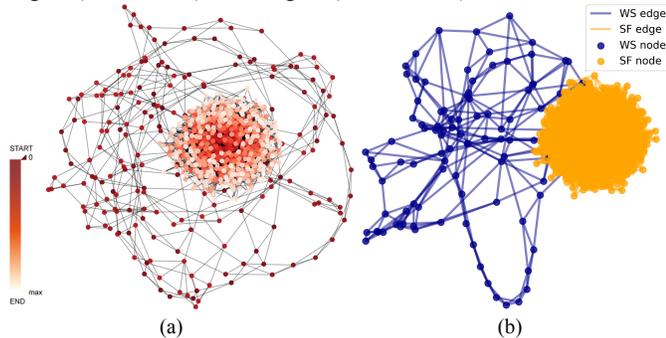

(a)        (b)

Fig. 7 Simulation of I network with N = 1,000, including 100 SW and 900 SF nodes. (a) Darker nodes are generated earlier, with the node color fading in the order in which the nodes are generated. The darker surrounding nodes in the diagram are SW networks formed earlier in the network growth process. The succeeding SF nodes generated are clustered toward the middle of the graph. (b) SF and SW nodes and links are colored with yellow and blue, respectively (recommendation: contrast with Fig. 1(a)).

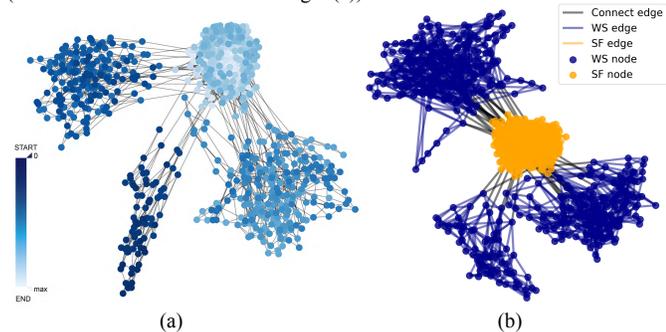

(a)        (b)

Fig. 8 Two further simulations of network II with N = 1,000 and 500 SW and SF nodes apiece. (a) Darker nodes correspond to earlier node generation. (b) SF and SW nodes and links are colored with yellow and blue, respectively (recommendation: contrast with Fig. 1(a)).

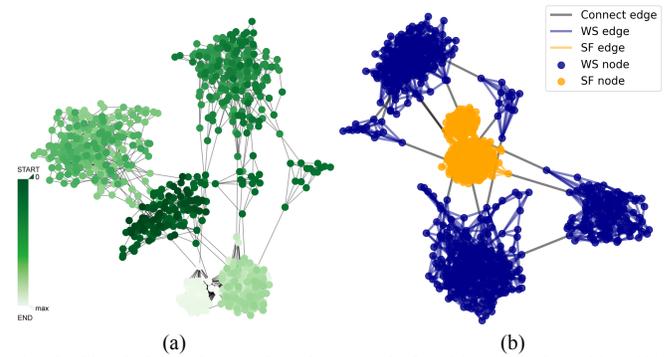

(a)        (b)

Fig. 9 Simulations of network III has a total of 1,000 nodes, also with a 60/40 SW/SF node breakdown. (a) Darker nodes correspond to earlier node generation. (b) SF and SW nodes and links are colored with yellow and blue, respectively (recommendation: contrast with Fig. 1(c)).

C) *Comparative Analysis of Three Hybrid Network Models*

We next compare how the three hybrid networks perform with the same mixtures of SW and SF nodes. Fig. 10a shows the distribution diagrams produced by the respective models with N = ten million nodes with a 50/50 SW/SF split. In Fig. 10b, the results are shown for the same networks with SW/SF ratios changed to 80/20. As a reference, Fig. 11 plots the results produced using (5) at the same 50/50 ratio used in Fig. 10(a). It is seen that the results in Fig. 11 follow the same trend as those produced by the networks II and III in Fig. 10(a) and are also similar to the trend produced by the network I. Although the mixing ratios used in Figs. 11 and 10(b) differ, the trends of the networks II and III in the latter figure are very similar to those produced using (5). Although the processes used to construct the respective models differ, all represent mixed networks with both SF and SW features. As noted in our previous mathematical discussion, the power-law-like distributions of networks with different nodes mixtures tend to converge experimental as the number of nodes increases. Although the model analyzed by Türker differs from that constructed in this paper, it also combines SW and SF network characteristics and produces a power-law-like distribution map similar to those produced here [28]. These results suggest that the degree distribution is most closely related to the network characteristics, with the method used to composed to construct the network having limited influence.

The results for the networks II and III in both Figs. 13(a) and (b) are quite similar up to $k = 10^2$, while those produced by the network I are close to the other two sets of results for $k \in [10^1, 10^2]$. Beyond $k = 10^{1.5}$, however, the respective curves begin to gradually diverge, primarily as a result of algorithmic differences. The network III has a double-layer, SF network construction, i.e., it generates subnets in the manner of an SF network. Based on the rules for multi-scale connection between subnets, SF subnets with larger average degrees are more easily selected as connections between subnets. Furthermore, selected nodes usually have larger degrees. For these reasons, the double-deck SF feature of the network III results in an exceedingly large node degree.

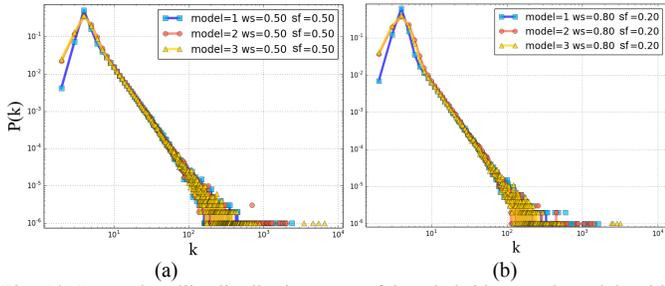

(a)　　　　　　　　　　　　　(b)

Fig. 10 Power-law-like distribution maps of three hybrid network models with (a) N= 10 million nodes and an SW/SF node feature split of 50/50 and (b) N = 10 million nodes and an SW/SF node feature split of 80/20.

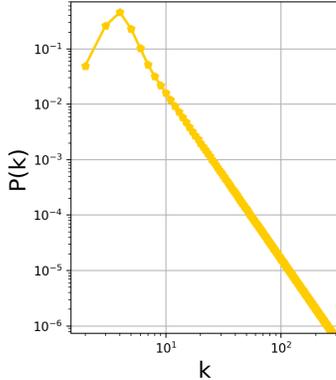

Fig. 11 Evolution of the output of (5) with: $P(k), k \in [2,1000], K = m = 4, a = 0.5, p = 0.3, N = 1,000$.

### D) Hybrid Propagation Model Simulation

We next simulate message propagation under the hybrid propagation model of an SF network. Fig. 12 shows the evolution of spreader density over time of networks with the following compositions: 80% SIS-type nodes, 15% SIR-type nodes, and 5% SIRS-type nodes (blue circles); 65% SIS nodes, 30% SIR nodes, and 5% SIRS nodes (red pentagons); and 50% SIS nodes, 45% SIR nodes, and 5% SIRS nodes (yellow squares). It is seen from a comparison of the three curves that increasing the proportion of SIS nodes in the hybrid propagation model increases the scope of the spread but delays the propagation peak.

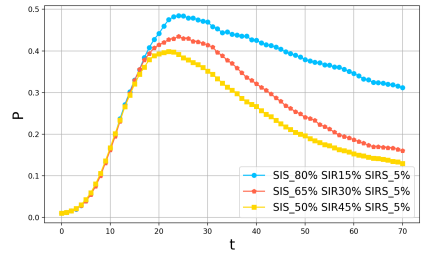

Fig. 12 Evolution of information spreader density over time under hybrid propagation models with SF networks and the results are averaged over 20 experiments.

### E) The Blockbuster Effect in The Hybrid Propagation Model Simulation

Finally, we look at message propagation in the three hybrid network types under the three propagation model conditions described above and under the presence of the blockbuster effect. Fig. 13(a) shows the evolution of spreader density with time in variants of the network I at three propagation model mixing ratios and a blockbuster trigger of 0.15. It is seen that the blockbuster is not triggered in the 50/45/5 SIS/SIR/SIRS network. Fig. 13(b) shows the evolution of spreader density over time in variants of the network II with the same propagation model mixtures and a blockbuster trigger of 0.07. Here as well, the blockbuster effect is not triggered in the network with a SIS/SIR/SIRS mixture of 50/45/5. Fig. 13(c) shows the results for the network III variants with the same propagation model mixtures and a blockbuster trigger of 0.048. Once again, no blockbuster effect is triggered in the network with a SIS/SIR/SIRS mixture of 50/45/5. All three simulations represent the average results of 20 experiments.

The consistent lack of blockbuster effect triggering in networks in which there is a SIS proportion of only 50% suggests that the hybrid propagation network in which nodes following the SIS model are not in the majority is less amenable to information propagation than the other two network types, resulting in a reduced scale of transmission in which propagation continues along implicit edge networks that are not triggered by the blockbuster effect.

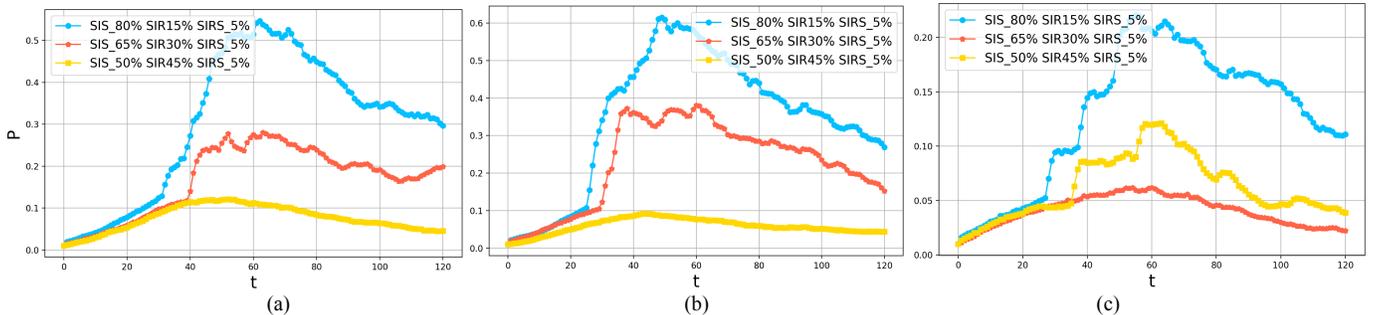

(a)　　　　　　　　　　　　　(b)　　　　　　　　　　　　　(c)

Fig. 13 (a). Evolution of information spreader density over time in variants of the I model with three propagation model mixtures. The blockbuster effect trigger is $\varphi = 0.1$ and the results are averaged over 20 experiments. (b). Evolution of information spreader density over time in variants of the II network with three propagation model mixtures. The blockbuster effect trigger is $\varphi = 0.07$ and the results are averaged over 20 experiments. (c). Evolution of information spreader density over time in variants of the III model with three propagation model mixtures. The blockbuster effect trigger is $\varphi = 0.048$ and the results are averaged over 20 experiments.

## V. COMPARATIVE ANALYSIS

To validate the applicability of the proposed theory to real social networks, we compare the simulation results to real data in the form of "K case" (a HFS case - Cancer discrimination case at a certain school in 2018/11) results crawled from an online web forum. Fig. 14(a) shows a comparison of the evolution of infection densities in the network I and the "K case" data. Based on the initial test, the trigger value of the blockbuster effect is 0.09. Fig. 14(b) shows the evolution of the blockbuster effect over time in the network I, with the reference showing the trigger value $\varphi = 0.09$. A comparison of Figs. 14(a) and 14(b) reveals that the blockbuster effect is triggered at $t = 43$ and $t = 45$ in the 80/15/5 (blue) and 66/30/5 (red) SIS/SIR/SIRS models, respectively. By contrast, in the 50/45/5 (yellow) model, the blockbuster effect is not triggered. Fig. 14(c) shows a comparison of the evolution of infection density in the network II with that in the "K case" data. Based on the second test, the trigger value of the blockbuster effect is 0.09. Fig. 14(d) shows the evolution of the blockbuster effect over time in the network II, with the reference line marking the trigger value $\varphi = 0.09$. Comparing Figs. 14(c) and 14(d) reveals results similar to those for the network I; that is, in the 80/15/5 (blue) and 65/30/5 (red) networks, the blockbuster effect is triggered at $t = 38$ and $t = 40$, respectively, while in the 50/45/5 (yellow) network, the blockbuster effect is not triggered.

The Euclidean metric can be used to define the degree of similarity between two curves, $\varsigma(t)$ and $\varrho(t)$, as follows:

$$similarity = \rho = \frac{\int_0^T \varsigma(t)dt - \int_0^T |\varsigma(t) - \varrho(t)|dt}{\int_0^T \varsigma(t)dt}. \quad (35)$$

The right-hand side of (35) gives the reciprocal of the sum of the distances between the two curves; as $\rho$ approaches one, the two curves become closer.

A comparison of the similarity between the results in Fig. 14(a) and the experimental data is shown in Fig. 14(e). It is seen that the transmission curve for the 65/30/5 SIS/SIR/SIRS network is most similar to that in the "K case," with a degree of similarity of $\rho \approx 0.9687$. By contrast, the 80/15/5 model has a degree of similarity to the "K case" of only $\rho \approx 0.3519$.

Fig. 14(e) also shows the similarities between the Fig. 14(c) results and the experimental data. In this case, the 55/45/5 network is most similar at $\rho \approx 0.3678$, while the 80/15/5 network has a similarity of only $\rho \approx 3678$.

To sum up, those comparisons validate the proposed model conform to the "K case".

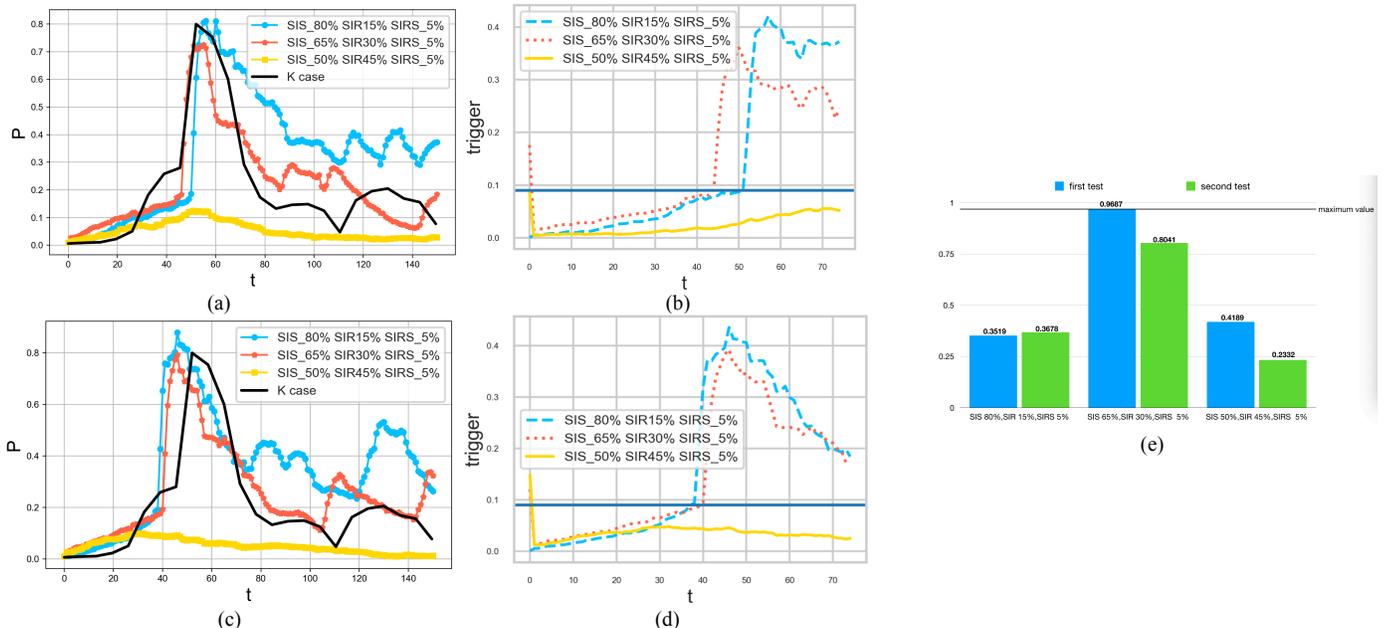

Fig. 14 (a). Comparison of evolution of first-test spreader density over time in the I networks with the "K case" data. Here, $\varphi = 0.09$, the result of the once experimental data comparison. (b) Evolution of the blockbuster effect over time during the first test. The reference line is the trigger value $\varphi = 0.09$, following the result of one experiment. (c) Comparison of evolution of second-test spreader density over time in the II networks with the "K case" data. Here, $\varphi = 0.09$, the result of the once experimental data comparison. (d) Evolution of the blockbuster effect with time in the second test. The reference line is the trigger value $\varphi = 0.09$, following the result of the one experiment. (e) Degree of similarity between the three hybrid model results over two experiments and the "K case" results.

## VI. CONCLUSION

In this paper, we developed and analyzed three hybrid network models with different mixtures of SF and SW network characteristics. The respective models were optimized and their generating algorithms were elaborated. We then investigated a hybrid propagation model comprising a mixture of SIS, SIR, and SIRS models. We also examined the blockbuster effect, a trigger mechanism that causes messages to spread between networks, and the concept of explicit-implicit edges. In addition, we investigated the degree distributions of the three hybrid network models and demonstrated that the network degree follows a power-law-like distribution that is closely related to the network characteristics but minimally affected by the network composition. We quantified this relation with a hybrid network model degree distribution formula. A comparison based on the Euclidean distance between simulation results and data obtained from a real HFS case from the Internet was used to validate the proposed model.

REFERENCES

[1] Lorsch, R. Jon, Nichols, and G. David, "Organizing Graduate Life Sciences Education around Nodes and Connections," *Cell,* vol. 146, no. 4, p. 506, 2011.
[2] A. L. Barabási, "Scale-Free Networks: A Decade and Beyond," *Science,* vol. 325, no. 5939, pp. 412-413, 2009.
[3] D. J. Watts, "A twenty-first century science," *Nature,* vol. 445, no. 7127, p. 489, 2007.
[4] P. S. Dodds, R. Muhamad, and D. J. Watts, "An Experimental Study of Search in Global Social Networks," *Science,* vol. 301, no. 5634, pp. 827-829, 2003.
[5] F. Schweitzer, G. Fagiolo, D. Sornette, F. Vegaredondo, A. Vespignani, and D. R. White, "Economic Networks: The New Challenges," *Science,* vol. 325, no. 5939, pp. 422-425, 2009.
[6] J. Bascompte, "Disentangling the Web of Life," *Science,* vol. 325, no. 5939, pp. 416-9, 2009.
[7] Vidal *et al.*, "Interactome Networks and Human Disease: Cell," *Cell,* vol. 144, no. 6, pp. 986-98, 2011.
[8] G. Yan *et al.*, "Network control principles predict neuron function in the Caenorhabditis elegans connectome," (in English), *Nature,* vol. 550, no. 7677, pp. 519-+, Oct 26 2017.
[9] R. Pastor-Satorras and A. Vespignani, *Evolution and Structure of the Internet: A Statistical Physics Approach*. Cambridge Univeristy Press, 2004, pp. 449 - 452.
[10] T. Berners-Lee, W. Hall, J. Hendler, N. Shadbolt, and D. J. Weitzner, "Creating a Science of the Web," *Science,* vol. 313, no. 5788, pp. 769-771, 2006.
[11] M. Y. Feng, H. Qu, Z. Yi, X. R. Xie, and J. Kurths, "Evolving Scale-Free Networks by Poisson Process: Modeling and Degree Distribution," (in English), *Ieee T Cybernetics,* vol. 46, no. 5, pp. 1144-1155, May 2016.
[12] P. Erdős and A. J. P. M. Rényi, "On Random Graphs I," vol. 4, pp. 3286-3291, 1959.
[13] P. Erdős and A. J. T. o. t. A. M. S. Rényi, "On the evolution of random graphs," vol. 286, no. 1, pp. 257-274, 2011.
[14] W. DJ and S. SH, "Collectivedynamics of 'small-world' networks," in *Nature*, 1998, pp. 440-442.
[15] A. L. Barabasi and R. Albert, "Albert, R.: Emergence of Scaling in Random Networks. Science 286, 509-512," vol. 286, no. 5439, pp. 509-512, 1999.
[16] P. Y. Chen, S. M. Cheng, and K. C. Chen, "Optimal Control of Epidemic Information Dissemination Over Networks," (in English), *Ieee T Cybernetics,* vol. 44, no. 12, pp. 2316-2328, Dec 2014.
[17] Y. Moreno, R. Pastor-Satorras, and A. Vespignani, "Epidemic outbreaks in complex heterogeneous networks," *The European Physical Journal B - Condensed Matter and Complex Systems,* vol. 26, no. 4, pp. 521-529, 2002.
[18] T. Harko, F. S. N. Lobo, and M. K. Mak, "Exact analytical solutions of the Susceptible-Infected-Recovered (SIR) epidemic model and of the SIR model with equal death and birth rates," (in English), *Appl Math Comput,* vol. 236, pp. 184-194, Jun 1 2014.
[19] I. Nåsell, "The Quasi-Stationary Distribution of the Closed Endemic SIS Model," *Advances in Applied Probability,* vol. 28, no. 3, pp. 895-932, 1996.
[20] V. M. Eguíluz and K. Klemm, "Epidemic Threshold in Structured Scale-Free Networks," *Physical Review Letters,* vol. 89, no. 10, p. 108701, 2002.
[21] H. Shi, Z. Duan, and G. Chen, "An SIS model with infective medium on complex networks," *Physica A Statistical Mechanics & Its Applications,* vol. 387, no. 8-9, pp. 2133-2144, 2008.
[22] J. Mena-Lorcat and H. W. J. J. o. M. B. Hethcote, "Dynamic models of infectious diseases as regulators of population sizes," vol. 30, no. 7, p. 693, 1992.
[23] C. H. Li, C. C. Tsai, and S. Y. Yang, "Analysis of epidemic spreading of an SIRS model in complex heterogeneous networks," (in English), *Commun Nonlinear Sci,* vol. 19, no. 4, pp. 1042-1054, Apr 2014.
[24] D. J. Daley and D. G. J. N. Kendall, "Epidemics and Rumours," vol. 204, no. 4963, pp. 1118-1118, 1964.
[25] F. Nian and S. Yao, "The epidemic spreading on the multi-relationships network," *Appl Math Comput,* vol. 339, pp. 866-873, 2018/12/15/ 2018.
[26] A. Chakraborty, H. Krichene, H. Inoue, and Y. Fujiwara, "Characterization of the community structure in a large-scale production network in Japan," (in English), *Physica A,* vol. 513, pp. 210-221, Jan 1 2019.
[27] M. F. Dai, Y. Zong, J. J. He, X. Q. Wang, Y. Sun, and W. Y. Su, "Two types of weight-dependent walks with a trap in weighted scale-free treelike networks," (in English), *Sci Rep-Uk,* vol. 8, Jan 24 2018.
[28] I. Turker, "Generating clustered scale-free networks using Poisson based localization of edges," (in English), *Physica A,* vol. 497, pp. 72-85, May 1 2018.
[29] K. Konstantin, V. M. Eguíluz, %J Physical Review E Statistical Nonlinear, and S. M. Physics, "Growing scale-free networks with small-world behavior," vol. 65, no. 2, p. 057102, 2002.
[30] K. Konstantin, V. M. Eguíluz, %J Physical Review E Statistical Nonlinear, and S. M. Physics, "Highly clustered scale-free networks," vol. 65, no. 3 Pt 2A, p. 036123, 2002.
[31] M. R. Sanatkar, W. N. White, B. Natarajan, C. M. Scoglio, and K. A. Garrett, "Epidemic Threshold of an SIS Model in Dynamic Switching Networks," (in English), *Ieee T Syst Man Cy-S,* vol. 46, no. 3, pp. 345-355, Mar 2016.
[32] Y. Wang, J. L. Ma, J. D. Cao, and L. Li, "Edge-based epidemic spreading in degree-correlated complex networks," (in English), *J Theor Biol,* vol. 454, pp. 164-181, Oct 7 2018.
[33] L. X. Yang, T. R. Zhang, X. F. Yang, Y. B. Wu, and Y. Y. Tang, "Effectiveness analysis of a mixed rumor-quelling strategy," (in English), *J Franklin I,* vol. 355, no. 16, pp. 8079-8105, Nov 2018.

[34] E. Lebensztayn and P. M. Rodriguez, "A connection between a system of random walks and rumor transmission," (in English), *Physica A,* vol. 392, no. 23, pp. 5793-5800, Dec 1 2013.

[35] N. Sumith, B. Annappa, and S. Bhattacharya, "RnSIR: A New model of Information Spread in Online Social Networks," (in English), *Proceedings of the 2016 Ieee Region 10 Conference (Tencon),* pp. 2224-2227, 2016.

[36] X. G. Zhang, C. H. Shan, Z. Jin, and H. P. Zhu, "Complex dynamics of epidemic models on adaptive networks," (in English), *J Differ Equations,* vol. 266, no. 1, pp. 803-832, Jan 5 2019.

[37] Z. J. Chen, W. Q. Tong, S. Kausar, and S. G. Zheng, "Two-Level-Mixing Rumor Propagation Model Based on Dynamic Trust Network," (in English), *Proceedings of 2016 International Conference on Audio, Language and Image Processing (Icalip),* pp. 658-662, 2016.

[38] F. Y. Wang *et al.*, "A Study of the Human Flesh Search Engine: Crowd-Powered Expansion of Online Knowledge," (in English), *Computer,* vol. 43, no. 8, pp. 45-53, Aug 2010.

[39] C. H. Chao, "Reconceptualizing the Mechanism of Internet Human Flesh Search: A Review of the Literature," (in English), *2011 International Conference on Advances in Social Networks Analysis and Mining (Asonam 2011),* pp. 650-655, 2011.

[40] B. Wang, B. A. Hou, Y. P. Yao, and L. B. Yan, "Human Flesh Search Model Incorporating Network Expansion and GOSSIP with Feedback," (in English), *Ieee Acm Dis Sim,* pp. 82-88, 2009.

[41] F. Y. Wang, D. Zeng, Q. P. Zhang, J. A. Hendler, and J. P. Cao, "The Chinese "Human Flesh" Web: the first decade and beyond," (in English), *Chinese Sci Bull,* vol. 59, no. 26, pp. 3352-3361, Sep 2014.

[42] Y. Zhang and H. Gao, "Human Flesh Search Engine and Online Privacy," (in English), *Sci Eng Ethics,* vol. 22, no. 2, pp. 601-604, Apr 2016.

[43] H. Zhu and B. Hu, "Agent based simulation on the process of human flesh search-From perspective of knowledge and emotion," (in English), *Physica A,* vol. 469, pp. 71-80, Mar 1 2017.

[44] L. Cheng, L. Zhang, and J. C. Wang, "A Study of Human Flesh Search with Epidemic Models," (in English), *Proceedings of the 3rd Annual Acm Web Science Conference, 2012,* pp. 67-73, 2012.

[45] Q. Li, C. G. Song, B. Wu, Y. P. Xiao, and B. Wang, "Social hotspot propagation dynamics model based on heterogeneous mean field and evolutionary games," (in English), *Physica A,* vol. 508, pp. 324-341, Oct 15 2018.

[46] S. Y. Tan, J. Wu, L. Y. Lu, M. J. Li, and X. Lu, "Efficient network disintegration under incomplete information: the comic effect of link prediction," (in English), *Sci Rep-Uk,* vol. 6, Mar 10 2016.

[47] K. Kandhway and J. Kuri, "Using Node Centrality and Optimal Control to Maximize Information Diffusion in Social Networks," (in English), *Ieee T Syst Man Cy-S,* vol. 47, no. 7, pp. 1099-1110, Jul 2017.

[48] Y. F. Wang, A. V. Vasilakos, J. H. Ma, and N. X. Xiong, "On Studying the Impact of Uncertainty on Behavior Diffusion in Social Networks," (in English), *Ieee T Syst Man Cy-S,* vol. 45, no. 2, pp. 185-197, Feb 2015.

[49] J. Gomez-Gardenes, L. Lotero, S. N. Taraskin, and F. J. Perez-Reche, "Explosive Contagion in Networks," (in English), *Sci Rep-Uk,* vol. 6, Jan 28 2016.

[50] F. Z. Nian and Z. K. Dang, "Message propagation in the network based on node credibility," (in English), *International Journal of Modern Physics B,* vol. 32, no. 9, Apr 10 2018.

[51] A. Barrat, M. J. T. E. P. J. B.-C. M. Weigt, and C. Systems, "On the properties of small-world network models," journal article vol. 13, no. 3, pp. 547-560, February 01 2000.

[52] B. M. J. I. J. s. a. C. Waxman, "Routing of Multipoint Connection," vol. 6, no. 9, pp. 1617-1622, 1988.

[53] A. Medina, I. Matta, and J. J. A. C. C. R. Byers, "On the Origin of Power Laws in Internet Topologies," vol. 30, no. 2, pp. 18-28, 2000.

[54] A. L. Barabási, R. Albert, and H. Jeong, "Mean-field theory for scale-free random networks," vol. 272, no. 1–2, pp. 173-187, 1999.

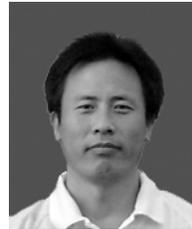

**Fuzhong Nian** received the B.S. degree of engineering from Northwest Normal University (department of Physics), LanZhou, China, in 1998, the M.S. degree in of engineering from Gansu University of Technology, Lanzhou, China, in 2004, the Ph.D. degree of engineering from Dalian University of Technology, Dalian, China, in 2011. He is interested in research at the intersection of mathematical modeling, network science, and control theory with application to biological, social and chaotic network.

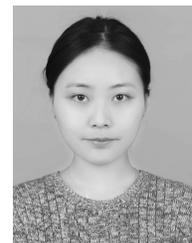

**Hongyuan Diao** received the B.S. degree in of engineering from Chongqing Normal University, Chongqing, China, in 2016, and is currently pursuing the M.S. degree in software engineering at Lanzhou University of Technology. Her main research interests include the modeling and analysis of complex networks, with applications in epidemic information and power grids.